# Subsymmetry-protected compact edge states

Ruoqi Cheng[1†], Domenico Bongiovanni[1,2†], Ziteng Wang[1†], Zhichan Hu[1], Liqin Tang[1,3]*, Daohong Song[1,3], Roberto Morandotti[2], Hrvoje Buljan[1,4], and Zhigang Chen[1,3]*

[1]*The MOE Key Laboratory of Weak-Light Nonlinear Photonics, TEDA Applied Physics Institute and School of Physics, Nankai University, Tianjin 300457, China*

[2]*INRS-EMT, 1650 Blvd. Lionel-Boulet, Varennes, Quebec J3X 1S2, Canada*

[3]*Collaborative Innovation Center of Extreme Optics, Shanxi University, Taiyuan, Shanxi 030006, China*

[4]*Department of Physics, Faculty of Science, University of Zagreb, Bijenička c. 32, Zagreb10000, Croatia*

[†]*These authors contributed equally to this work*

*e-mail: tanya@nankai.edu.cn, zgchen@nankai.edu.cn

**Sub-symmetry (SubSy) protected topological states represent a concept that goes beyond the conventional framework of symmetry-protected topological (SPT) phases, demonstrating that topological boundary states can remain robust even when the pertinent symmetry holds only in a subset of Hilbert space. Typical SPT and SubSy boundary states decay exponentially into the bulk, which means they are not confined in just few lattice sites close to the boundary. Here, we introduce topologically compact edge states protected by SubSy, featuring extreme two-site localization at boundaries of a lattice, without any decay into the bulk. The compactness arises from local destructive interference at the boundary, while topological protection is ensured by SubSy, characterized by quantized winding numbers. Experimentally, we observe compact edge states in laser-written photonic lattices with engineered rhombic-like unit cells, confirming their robustness against perturbations under both chiral symmetry and SubSy conditions. Our results highlight the potential of SubSy protection for achieving topological confinement of light, paving the way for applications in compact waveguides, lasers, and high-sensitivity photonic sensors.**

**Introduction:**

Over the past two decades, topological insulators (TIs) [1-5] have emerged as a groundbreaking class of materials by establishing the universal paradigm of bulk-boundary correspondence (BBC), which links nontrivial bulk invariants to robust boundary states. Initially introduced in condensed matter physics, TIs have rapidly sparked significant research interests across a variety of multidisciplinary fields, including electric circuits [6, 7], ultracold atoms [8-10], acoustics [11, 12], and photonics [3, 5, 13-16], fundamentally reshaping our understanding of the phase of matter and wave transport dynamics. Topologically protected boundary states, renowned for their exceptional robustness against disorder and fabrication imperfections, represent a hallmark feature of topological phases. This property has been extensively explored and highly tested for disruptive technologies, ranging from topological lasers [17-20] to fault-tolerant quantum computing [21]. Fueled by the promise of robust, scalable, and defect-immune light control, topological photonics has rapidly emerged as a leading frontier, poised to drive a new wave of topological phenomena in photonics, acoustics and other platforms with far-reaching impact in the coming decade [22, 23].

A particularly important class of topological phases are symmetry-protected topological (SPT) phases, where boundary states remain robust to perturbations preserving specific symmetry (or symmetries), but the topology is lost once the symmetry is broken [4, 5]. Such a symmetry dependence brings about opportunities for engineering topological responses but also fundamental limitations for practical implementations. In most SPT phases such as for topological crystalline insulators, the manifestation of topologically trivial or nontrivial boundary states is often tied to a lattice dimerization or coupling parameter [24-27]. Typically, the amplitude distributions of topologically nontrivial boundary states display an exponential decay from the boundary regions towards the bulk, developing a characteristic phase relation. Therefore, such boundary states are not strictly confined to a small number of lattice sites and thus are not highly localized. On the other hand, strong spatial localization rather than extended Bloch waves can be realized in flat-band lattices such as in Kagome [28, 29], Lieb [30-32], and decorated honeycomb [33-35] geometries. These models support compact localized states (CLSs) that originate from macroscopic degeneracies of Bloch modes and destructive interference enforced by lattice geometry [36-39]. The absence of group velocity across the Brillouin zone leads to localized modes confined to a few lattice sites, without the need for external disorder or nonlinear effects. In this context, realization of Aharonov-Bohm cages [40-42] and compact bulk or boundary light localizations [32, 43-45] has been reported by leveraging the unique properties of flat-band modes. However, light localization relies on flat-band dispersion does not necessarily possess topological protection, making it susceptible to perturbations. Likewise, defect lattices of mirror-symmetry can support symmetry-protected bound states in the continuum (BICs) [46, 47], which remain localized due to destructive interference even when their energy lies within the continuum band of extended modes. Although compactly localized, those states do not show robustness under perturbations as they are not topologically protected [48, 49].

Recent advances on SPT phases have redefined protection criterions through the introduction of sub-symmetry (SubSy) - a framework in which symmetry operators act exclusively within designated Hilbert subspaces [50-53]. Within this framework, topological boundary states belonging to the subspace are protected as long as perturbations do not break the corresponding SubSy, without requiring the full global symmetry of the entire system. This opens a new pathway to protect topological edge states beyond stringent global-symmetry constraints.

In this work, we exploit the concept of SubSy protection to demonstrate a new class of topologically compact edge states, analogous to dipole-like modes that are highly localized at the lattice boundary. The underlying model is a quasi-one-dimensional (1D) rhombic-like lattice with nearest-neighbor (NN) hopping uniform across the lattice, supplemented with next-nearest-neighbor (NNN) hopping, which breaks the global chiral symmetry (CS) but retains the lattice SubSy. Unlike flat-band CLSs, these compact edge states do not rely on a flat band. Instead, they are characterized by pertinent SubSy with a corresponding topologically nontrivial winding number. Thus, the boundary states exhibit simultaneously the compactness and topological robustness, expanding the design landscape for highly localized topological boundary states. In the experiment, SubSy-protected compact edge states are realized in laser-written photonic lattices with desired lattice configurations, and compared with the SubSy-broken case. Our results constitute an effective synergy between local destructive interference and topological SubSy protection, opening an avenue for applications in compact waveguides and high-sensitivity sensors.

**Theory:**

Schematic illustration in the inset of Fig. 1(a) depicts a quasi-1D periodic rhombic-like lattice, where each unit cell (delineated by the dashed square) consists of four sublattices labeled A, B, C, and D. The model is partitioned into two subsystems: the AB-subsystem (marked in red) and the CD-subsystem (marked in green). The coupling coefficient $t$ corresponds to all NN interactions across the entire photonic lattice, regardless of whether the coupling occurs within the unit cell (intracell hopping) or between different unit cells (intercell hopping). The theoretical model comprises an NNN coupling, denoted as $t_1$ (see the inset of Fig. 1(a)). This $t_1$ coupling represents intracell coupling between sites A and B, as well as intercell coupling between sites C and D. When this NNN coupling is present, the global CS is broken, while the SubSy is preserved. The momentum space Bloch Hamiltonian for this lattice is

$$H(k) = \begin{pmatrix} 0 & t_1 & t & t+te^{-ik} \\ t_1 & 0 & t & t \\ t & t & 0 & t_1 e^{-ik} \\ t+te^{ik} & t & t_1 e^{ik} & 0 \end{pmatrix}, \quad (1)$$

where $k$ is the Bloch momentum.

To investigate the topological properties of the system, let us first discuss the case $t_1 = 0$, for which the Bloch Hamiltonian $H$ is of the form

$$H(k) = \begin{pmatrix} 0 & Q(k)^\dagger \\ Q(k) & 0 \end{pmatrix}, \quad (2)$$

with $Q(k)$ denoting the lower-left block. The Hamiltonian $H$ exhibits CS, since we have $\Sigma H(k) \Sigma^{-1} = -H(k)$, where $\Sigma$ is the CS operator:

$$\Sigma = \begin{pmatrix} I_2 & 0 \\ 0 & -I_2 \end{pmatrix}, \quad (3)$$

with $I_2$ denoting a $2 \times 2$ identity matrix. By diagonalizing the Bloch Hamiltonian, we obtain the energy band structure [Fig. 1(b1)], featuring a band gap near zero energy. As in a typical 1D chiral-symmetric system, the topological phase can be characterized by the winding number [54, 55]

$$W = \frac{1}{2\pi} \int_0^{2\pi} dk \frac{d\theta(k)}{dk}, \quad (4)$$

where $\theta(k)$ is defined by $\det Q(k) = |\det Q(k)| e^{i\theta(k)}$. The winding number counts the number of times that $\det Q(k)$ encircles the origin in the complex plane as $k$ varies across the Brillouin zone.

In this case of quasi-1D rhombic-like lattice, with uniform hopping $t$ and $t_1 = 0$, we obtain the winding number $w = 1$ (inset, Fig. 1(b1)). A nonzero winding number indicates the presence of a pair of topological edge states at zero-energy $\beta = 0$. To verify their existence, we numerically solve the eigenvalue problem for a finite lattice consisting of 20 unit cells [Fig. 1(b2)]. The chiral symmetric lattice supports two eigenvalues in the gap pinned to zero energy, which correspond to two topological edge states. The spatial distributions of these edge states are plotted as insets in Fig. 1(b2). Strikingly, these edge states have non-zero amplitudes only at two boundary sites – defining them as "compact edge states". Such intrinsic compactness departs from conventional topological edge states (e.g., those in the SSH [24] and trimer [26] models), whose wavefunction amplitudes decay exponentially from the boundary towards the bulk.

To quantify the degree of localization of compact edge states, a typical measure is the inverse participation ratio (IPR), defined as IPR $= \sum|\psi_n|^4 / (\sum|\psi_n|^2)^2$. The IPR value ranges from 0 to 1, with a small value indicating an extended state and a large value signifying strong localization. Clearly, one can see that both topological edge states exhibit an IPR of 0.5 (green dots in Fig. 1(b3)). This value explicitly confirms their compact nature: wavefunction amplitudes are strictly confined to two boundary lattice sites without any leakage into other lattice sites [56].

Unlike conventional 1D topological systems (e.g., the SSH model and its variations) which require a dimerization (i.e., "dimers" with different coupling strengths along the chain) for nontrivial phases, the topological edge states uncovered here emerge in uniformly coupled lattices with identical coupling strengths. These states exhibit several defining attributes: compact localization to exactly two boundary sites with opposite phase in wavefunction amplitudes, and confinement within a single subsystem (i.e., AB- or CD-subsystem). Their compactness is a result of destructive interference of boundary states. Our system contains no flat bands, and thus localization differs in nature from that in flat-band models, where it is mediated by band degeneracy with zero momentum-dependent energy dispersion [45, 57, 58]. Our results thus highlight an effective interplay between local boundary destructive interference and nonlocal bulk topology.

Next, we analyze the model when $t_1 \neq 0$. In this case, the global CS is broken, yet an SubSy still maintains and topologically protects the left-edge state, whereas the right-edge state loses its compactness and has no topological protection. Theoretically we validify this statement by linearly combining the A and B sublattices and defining the two new sublattices

$$|A'\rangle = (|A\rangle + |B\rangle)/\sqrt{2}, \tag{5}$$

$$|B'\rangle = (|A\rangle - |B\rangle)/\sqrt{2}. \tag{6}$$

The procedure is equivalent to carrying out a unitary transformation $H'(k) = \Gamma H(k)\Gamma^{-1}$, where

$$\Gamma = \begin{pmatrix} 1/\sqrt{2} & 1/\sqrt{2} & 0 & 0 \\ 1/\sqrt{2} & -1/\sqrt{2} & 0 & 0 \\ 0 & 0 & 1 & 0 \\ 0 & 0 & 0 & 1 \end{pmatrix}, \tag{7}$$

and the transformed Hamiltonian $H'(k)$ reads as

$$H'(k) = \begin{pmatrix} 2t_1 & 0 & \sqrt{2}t & \frac{(2+e^{-ik})t}{\sqrt{2}} \\ 0 & 0 & 0 & \frac{e^{-ik}t}{\sqrt{2}} \\ \sqrt{2}t & 0 & t_1 & e^{-ik}t_1 \\ \frac{(2+e^{ik})t}{\sqrt{2}} & \frac{e^{ik}t}{\sqrt{2}} & e^{ik}t_1 & t_1 \end{pmatrix} - t_1 I_4, \tag{8}$$

where $I_4$ denotes a 4 × 4 identity matrix. In the transformed model described by $H'(k)$, $B'$-SubSy is preserved: $\Sigma_{B'}^{\dagger}(H'(k) + t_1 I_4)\Sigma_{B'}P_{B'} = -(H'(k) + t_1 I_4)P_{B'}$, where $P_{B'}$ is the projection operator on the $B'$ sublattice and $\Sigma_{B'} = 2P_{B'} - I_4$. From Eq. (8), the $B'$ sublattice only couples to the D sublattice via an intercell hopping of $t/\sqrt{2}$, with no intracell hopping. The scenario is quite analogous to a topologically nontrivial 1D SSH model in the limiting dimerization regime where the intracell hopping is zero. This limiting SSH model supports a compact edge state residing only on one site, which can be directly mapped into a compact edge state supported by the transformed Hamiltonian $H'(k)$: $|\psi'\rangle = [0,1,0,\ldots,0]^T$ with an energy $\beta = -t_1$. Consequently, this edge state only resides in the $B'$ sublattice and is protected by $B'$-SubSy.

It is important to mention that only the edge state of the limiting SSH model can be mapped to the $H'(k)$, being confined to one sublattice. In contrast, bulk states cannot be mapped because they do not reside on one sublattice. Since the $B'$ sublattice corresponds to a linear combination of the original A and B sublattices, the actual left-edge state thus populates only A and B sites, and can be analytically expressed as $|\psi\rangle = \Gamma^{-1}|\psi'\rangle = [1,-1,0,\ldots,0]^T/\sqrt{2}$ with an energy $\beta = -t_1$. As illustrative examples, Fig. 1(c1) plots the $k$-space band structure for $t = 1$ and $t_1 = 0.4$. For the finite-size lattice, the real-space eigenvalue spectrum is illustrated in Fig. 1(c2) along with the mode distributions of edge states. The left-edge state remains within the band gap with an IPR of 0.5 (see Fig. 1(c3)), indicating its compactness. In contrast, the right-edge state nearly merges into the bulk band – evidenced by its diminished IPR [Fig. 1(c3)], losing both topological robustness and spatial localization due to no SubSy protection in this case. Furthermore, we numerically calculate the real-space eigenvalues as a function of the coupling

$t_1$ (see Fig. 1(a)). The left-edge state (labeled by the green line in Fig. 1(a)) persists as $t_1$ varies from -1 to 1 with an energy $\beta = -t_1$. These numerical results agree well with our theoretical predictions, confirming that when $t_1 \neq 0$, the left-edge state remains protected by $B'$-SubSy and retains its compactness, whereas the right-edge state loses its compact feature and populates onto both AB and CD sublattices (see the top inset in Fig. 1(c2)).

**Experiment**

To observe the predicted compact edge states in experiment, we create two distinct quasi-1D photonic lattices – one preserves the global CS ($t_1 = 0$), and the other only preserves $B'$-SubSy ($t_1 \neq 0$). Lattice structures with a spacing of 35 μm are established in a photorefractive crystal (SBN:61) via a laser-writing technique [32, 59, 60]. To accurately implement the theoretical model depicted in Fig. 1(a), we employ a staggered lattice structure by flipping the adjacent unit cells (flipping B and C sites along the y-axis while keeping A and D sites unchanged) [Fig. 2(a1)]. This configuration ensures that intercell interactions are restricted solely to A and D sites in this system. Figure 2 presents both experimental and numerical results for the chiral-symmetric lattice where the coupling coefficient $t_1$ is appropriately set to zero (see the theoretical model in Fig. 1(a) for reference). In our experimental setup, the couplings are precisely controlled by their inter-waveguide distances. We achieve variation in $t_1$ through adjusting the angle $\theta$, as defined in the inset of Fig. 2(a1). Specifically, when $\theta = 90°$, $t_1$ is approximately 0. Figure 2(a1) shows the experimental CS-preserving lattice, corresponding to the Hamiltonian $H$ in Eq. (1) when $t_1 = 0$. The probe beams are modulated into a dipole-shaped pattern carrying an appropriate phase relationship: an out-of-phase configuration [Fig. 2(a2)], which matches the mode distribution of topological edge states, and an in-phase configuration [Fig. 2(a3)] for comparison. To excite the compact edge state, the out-of-phase probe beam is launched into the A and B sublattices at the left edge. As shown in Fig. 2(b1), after 20 mm-long propagation, the output intensity remains perfectly localized on the excited waveguides while preserving the out-of-phase relationship (see inset in Fig. 2(b1)). In contrast, for the in-phase excitation, light couples into the C and D sublattices, as observed in Fig. 2(c1). This occurs because in-phase excitation does not match the mode of the edge states. Bulk excitations, whether in out-of-phase [Fig. 2(d1)] or in-phase [Fig. 2(e1)] configurations, cannot maintain confinement. In addition, corresponding numerical simulations [Figs. 2(b2)-2(e2)] carried out with parameters from the experiment closely match the experimental results [Figs. 2(b1)-2(e1)] under the same propagation length of 20 mm. For a better comparison of the dynamics, we also performed numerical simulations to a much longer propagation distance of 100 mm [Figs.

2(b3)-2(e3)]. After long propagation, the out-of-phase beam still remains localized at the A and B sites of the left edge [Fig. 2(b3)], but the beam spreads into nearby waveguides for other excitation conditions [Figs. 2(c3)-2(e3)]. These numerical results further confirm the existence of topological compact edge states.

Next, we perform a series of experiments to demonstrate the preservation and breakdown of compact edge states in the $B'$-SubSy lattice, as shown in Fig. 3. By adjusting the angle $\theta$, the NNN hopping $t_1$ can be introduced. When the angle $\theta$ exceeds 90 degrees, $t_1$ is nonzero, which in turn breaks CS although preserving $B'$-SubSy, as illustrated in Figs. 3(a1) and 3(a2). After 20 mm of propagation, the probe beam that matches the out-of-phase structure of the topological compact edge state remains intact upon launching into the A and B sites at the left boundary [Fig. 3(b1)]. In contrast, when the same excitation occurs on the right edge, confinement on the C and D sites is not observed. Instead, due to the absence of SubSy protection, light spreads to other sublattice sites that belong to the nearby unit cell, as shown in Fig. 3(b2). Under the in-phase excitation condition [Figs. 3(c1) and 3(c2)], the output intensities after 20mm propagation exhibit more light spreading compared to the out-of-phase excitation. Numerical simulations with out-of-phase inputs, presented in Figs. 3(d1) and 3(d2), show good agreement with the experimental results [Figs. 3(b1) and 3(b2)]. To highlight the contrast more clearly, we present simulation results with a propagation length of 100mm, as shown in Figs 3(e1) and 3(e2). We can see that the left edge state remains well-localized even after long-distance propagation, whereas on the right edge the probe beam spreads significantly into the bulk, further confirming the need of SubSy protection for realization of compact edge states.

**Discussion**

Robustness against perturbations that preserve the underlying symmetries is a defining characteristic of any topological boundary state associated with an SPT phase. In this regard, it is important to test the stability of topological compact edge states under perturbations that respect the relevant symmetry (symmetries). Let us first consider robustness of the compact edge state under perturbations of the coupling parameters $t$ and $t_1$ that respect the lattice symmetry. Such a perturbation leaves the system within the phase diagram studied above: for $t_1 = 0$ the compact edge state is preserved under CS, while for $t_1 \neq 0$ the SubSy is responsible for the robustness of the compact edge state.

However, random perturbations that do not preserve the lattice symmetry are more disruptive. To study these perturbations, we randomly vary the coupling terms in the associated real-space Hamiltonian [Fig. 1(a)]. Every hopping term between lattice sites $n$ and $m$ incorporates a disorder component: $t \to t + d \times \xi_{nm}$, $t_1 \to t_1 + d \times \xi_{nm}$, where $d$ is the disorder strength, and $\xi_{nm}$ is a random value in the interval $[-1,1]$ (the couplings that are not present in the unperturbed case remain zero). Figure 4 presents numerical results of perturbation tests performed on both chiral symmetric ($t_1 = 0$) and SubSy ($t_1 \neq 0$) models. The upper row in Fig. 4(a1-a4) illustrates the eigenvalues of the system as a function of the perturbation strength, whereas the lower row in Fig. 4(b1-b4) illustrates the IPR of each state. Without loss of generality, we take the average of the results after performing calculations on 100 sets of random configurations.

For the chiral symmetric lattice with hopping parameters $t = 1$ and $t_1 = 0$, the disorder is introduced only on $t$ to preserve CS, see Figs. 4(a1) and 4(b1). From the perturbed band structure in Fig. 4(a1), the eigenvalues of both edge states (highlighted by the red line) remain fixed at zero energy within the energy gap, as long as before the disorder strength becomes sufficiently large to introduce band-gap closure. Consistently, the IPR of edge states gradually decreases as the disorder strength increases, eventually merging with the IPR of bulk states once the band gap closes [Fig. 4(b1)].

For the SubSy preserving lattice with $t = 1$ and $t_1 = 0.4$, the disorder is added to both $t$ and $t_1$ hopping parameters, see Figs. 4(a2) and 4(b2). The perturbed band structure of the SubSy lattice shows the left-edge state being protected in the band gap while the right-edge state loses protection (see Fig. 4(a2) inset): the eigenvalues of the left-edge state (red lines, Fig. 4(a2)) remain within the energy gap unless the strength of the disorder significantly increases to merge it into the bulk bands. In terms of the IPR [Fig. 4(b2)], the left-edge state exhibits similar behavior to the chiral symmetric case [Fig. 4(b1)] whereas the IPR of the right-edge states (yellow dots) decreases much more due to the lack of protection.

Thus, for the CS and SubSy preserving lattices, we find that the topological edge state is protected and localized under perturbations that respect the pertinent symmetry (CS or SubSy); however, the IPR calculations show that compactness is not fully preserved under such random perturbations that break the lattice symmetry.

Next, we may ask the following question: can we have random disordered perturbations that keep also the compactness intact? We find that the additional condition required to preserve the compactness of edge states is: $\frac{t_{AC}}{t_{BC}} = \frac{t_{AD}}{t_{BD}}$, where $t_{\alpha\beta}$ represents the intercell hopping between sublattices $\alpha$ and $\beta$ (here $\alpha$ and $\beta$ stands for any of the sublattice sites, A, B, C or D). In Figs. 4(a3), 4(b3), 4(a4) and 4(b4), we repeat numerical tests when the compactness condition is additionally satisfied apart from the relevant symmetry. For the chiral symmetric lattice, the compactness condition is $\frac{t_{AC}}{t_{BC}} = \frac{t_{AD}}{t_{BD}}$, see Figs. 4(a3) and 4(b3). The perturbed band structure and the IPR show that the two edge states not only remain robust within the band gap but also retain their compactness as examined by the large IPR. Note that a large IPR implies the state is still localized on two sites but with unequal amplitudes. For the $B'$-SubSy lattice, the compactness condition is $\frac{t_{AC}}{t_{BC}} = \frac{t_{AD}}{t_{BD}} = 1$, see Figs. 4(a4) and 4(b4). The band structure and IPR reveal a protected left-edge state whose IPR always remains 0.5, indicating compact localization with equal amplitudes on the two sites. Additionally, the insets in Fig. 4(a1-a4) show clearly different profiles of protected compact edge states and non-protected states when the disorder strength is 0.2.

**Conclusion**

In conclusion, we have theoretically and experimentally demonstrated the existence of topologically compact edge states in a 1D photonic rhombic-like lattice. Such edge states are topologically protected by the chiral symmetry or SubSy. Different from exponentially decaying topological states, compact edge states exhibit strict confinement to the two boundary sites. This unprecedented localization precision unlocks transformative applications in ultra-compact topological nanocavities and single-emitter sensors. Our findings not only reveal a fundamentally new form of boundary physics in topological photonics but also redefine design principles for symmetry-protected topological matter, offering broad implications for quantum coherence control and topological engineering across diverse physical platforms.

**Acknowledgments**
This research is supported by the National Key R&D Program of China under Grant No. 2022YFA1404800, the National Natural Science Foundation (W2541003, 12134006, 12374309, 11922408, 124B2078, 12504385). H.B. acknowledges support from the project "Implementation of cutting-edge research and its application as part of the Scientific Center of Excellence for Quantum and Complex Systems, and Representations of Lie Algebras", Grant No. PK.1.1.10.0004, co-financed by the European Union through the European Regional Development Fund - Competitiveness and Cohesion Programme 2021-2027. Z.H. acknowledges the support of China Postdoctoral Science Foundation (Grant Nos. BX20240174). D.B. acknowledges support from the Ministry of Human Resources and Social Security of China (Grant WGXZ2023110). R.M. acknowledges support from NSERC Discovery and the CRC program in Canada.

**Conflict of interest**

The authors declare no conflicts of interest and no competing financial interests.

**Contributions**

All authors participated in and contributed to this work. Z.C. and H.B. supervised the project.

Correspondence and requests for materials should be addressed to Z.C. or L.T.

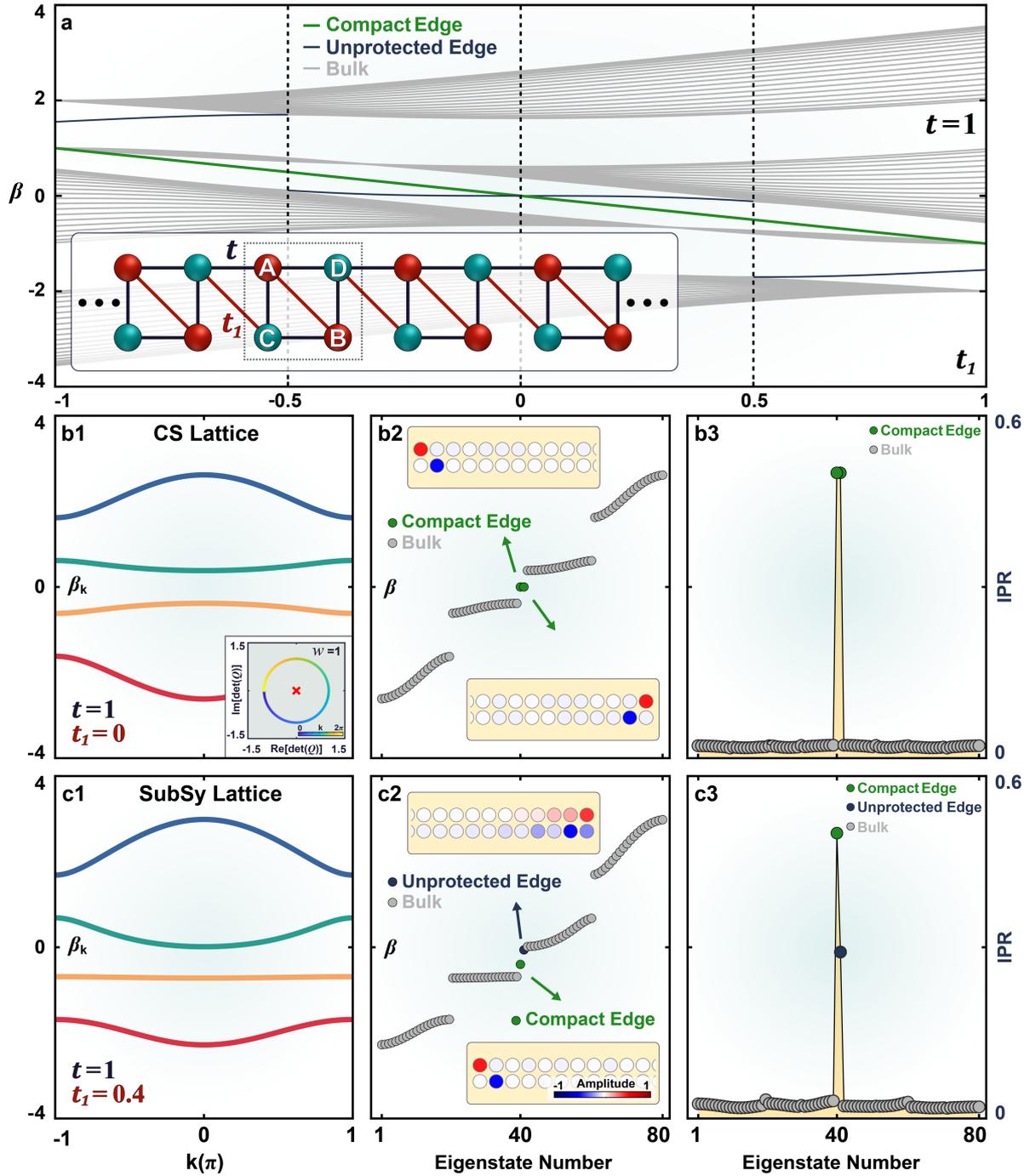

**Fig. 1 Theoretical analysis of topological compact edge states in chiral symmetry and SubSy lattices.** Inset in (a) shows a schematic illustration of a quais-1D rhombic-like lattice, where $t$ accounts for NN hopping among adjacent sites and $t_1$ for the NNN hopping, and each unit cell contains four sublattices labeled A, B, C, and D. (a) Eigenvalue spectral evolution as a function of $t_1$ in real space calculated for $t = 1$, where the green line highlights compact edge states. (b1, c1) $k$-space band structure for $t = 1$ of (b1) chiral symmetric ($t_1 = 0$) and (c1) SubSy ($t_1 = 0.4$) lattices, respectively. Note that the seemingly flat band indicated in yellow in (c1) is not, in fact, a true flat band. Inset in (b1) shows the winding number calculation. (b2, c2) Corresponding eigenvalue spectrum for a finite-sized lattice consisting of 20 unit cells. Insets

display the eigenmode distributions corresponding to the left- and right-edge states. (b3, c3) The inverse participation ratios (IPR) of each eigenstate for (b3) the chiral symmetric lattice ($t_1 = 0$) and (c3) SubSy lattice ($t_1 = 0.4$).

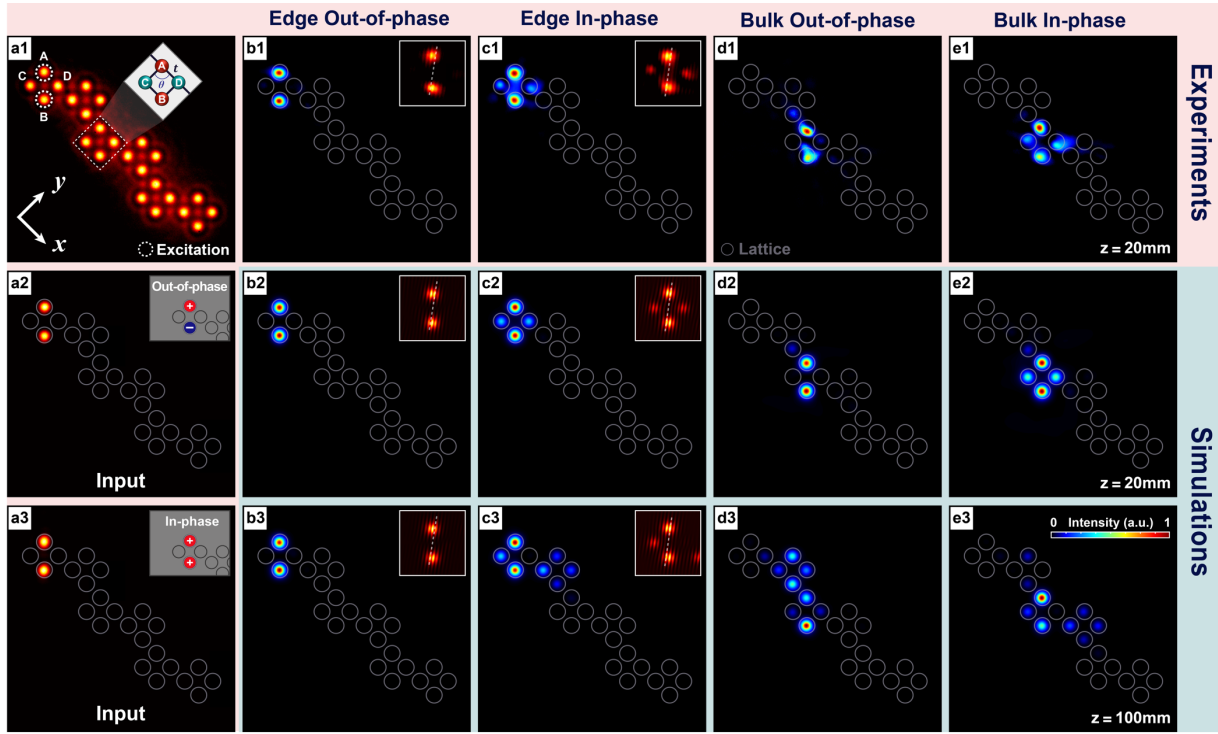

**Fig. 2 Experimental observations of topological compact edge states in a chiral symmetry lattice.** (a1) Staggered chiral symmetric photonic lattice. (a2, a3) Input intensity pattern of probe dipole-shaped beam with out-of-phase (a2) and in-phase (a3) relationship, respectively. (b1-e1) Output intensity pattern of probe beams after 20 mm-long propagation corresponding to (b1) edge out-of-phase, (c1) edge in-phase, (d1) bulk out-of-phase, and (e1) bulk in-phase excitations, respectively. (b2-e2, b3-e3) Numerical simulations with the same parameters as experiments performed over (b2-e2) 20 mm and (b3-e3) 100 mm propagation distance. Insets in (b1-b3, c1-c3) show output interferograms.

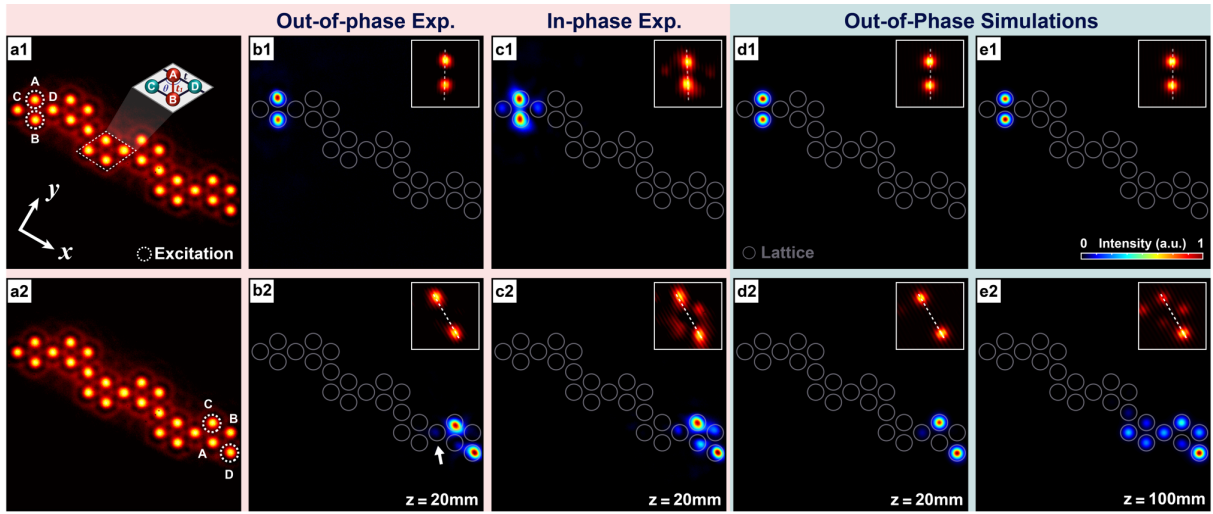

**Fig. 3 Experimental observations of topological compact state in a stretched SubSy lattice.** (a1, a2) SubSy photonic lattices obtained by appropriately adjusting the chiral-symmetric structure shown in Fig. 2(a1) (achieved by stretching the angle $\theta$). White dashed circles highlight probe-beam excitations: (a1) exciting the preserved compact left-edge states within the AB sublattices, and (a2) exciting the broken right-edge states within the CD sublattices. (b1, b2) Experimental output intensities after 20 mm-long propagation under out-of-phase excitations for (b1) left- and (b2) right-edge excitations. (c1, c2) Experimental output intensities after 20-mm propagation under in-phase excitations for left and right edges, respectively. (d1-d2, e1-e2) The numerical simulations under out-of-phase excitations for left and right edges at (d1, d2) $z = 20mm$ and (e1, e2) $z = 100mm$. The top-right insets in (b1-e1, b2-e2) show the output interferograms.

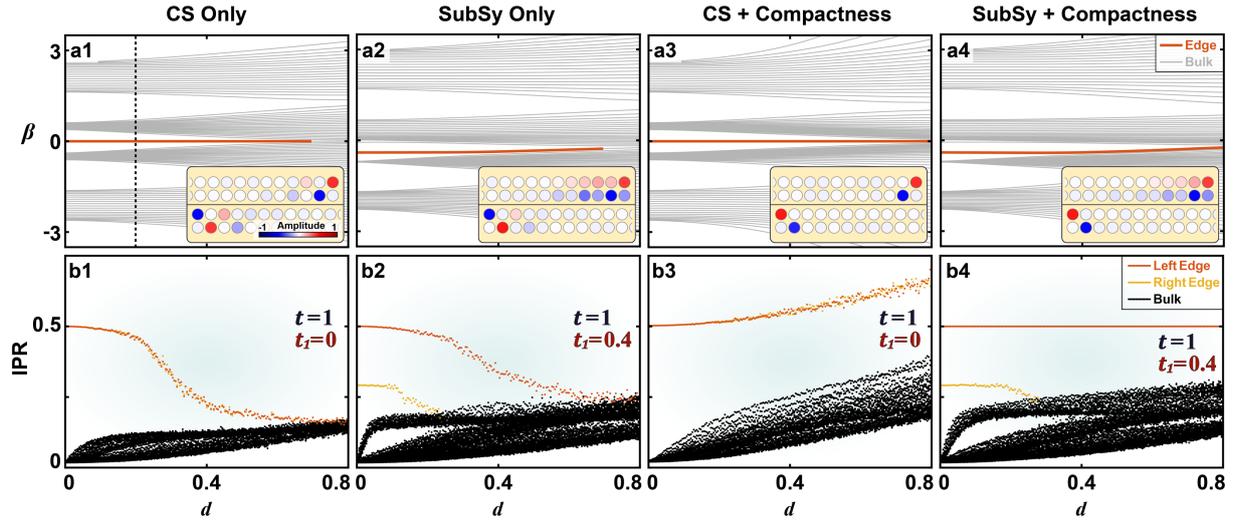

**Fig. 4 Robustness test for topological compact edge states.** (a1-a4) Perturbed energy spectral evolution for different disorder strength corresponding to a perturbed lattice preserving (a1) chiral symmetry (CS) only, (a2) $B'$-SubSy only, the transformed sublattice $B'$ defined in Eq. (6) of the main text, (a3) CS and compactness condition, and (a4) $B'$-SubSy and compactness condition. (b1-b4) Inverse participation ratio (IPR) as a function of $d$ under the same conditions as in (a1-a4), where orange and yellow dotted lines denote the left- and right-edge states, respectively. Insets in (a1-a4) show mode distributions of left- (lower panel) and right- (upper panel) edge states for $d = 0.2$.